\documentclass[nologo,11pt,a4paper]{ETHpaper} 
\usepackage{microtype}
\usepackage{color, colortbl}
\usepackage{url}
\usepackage{hyperref} 
\usepackage{mathtools}
\usepackage{graphicx} 
\usepackage{multirow}   
\usepackage{subfigure} 
         
\definecolor{Gray}{rgb}{0.9,0.9,0.9}
   
\begin{document} 

\title{A Quantitative Study of Social Organisation in Open Source Software Communities}
 
\titlealternative{A Quantitative Study of Social Organisation in Open Source Software Communities}

\author{Marcelo Serrano Zanetti, Emre Sarig\"ol, Ingo Scholtes, Claudio Juan Tessone, Frank Schweitzer} 

\authoralternative{Marcelo Serrano Zanetti, Emre Sarig\"ol, Ingo Scholtes, Claudio Juan Tessone, Frank Schweitzer}

\address{Chair of Systems Design, ETH Zurich,\\ Zurich, Switzerland,
  \texttt{mzanetti@ethz.ch}} 

\reference{ICCSW 2012, pp. 116--122, DOI: 10.4230/OASIcs.ICCSW.2012.116}

\www{\url{http://www.sg.ethz.ch}} 

\makeframing
  
\maketitle

\begin{abstract}
The success of open source projects crucially depends on the voluntary contributions of a sufficiently large community of users.
Apart from the mere size of the community, interesting questions
arise when looking at the \emph{evolution of structural features} of collaborations between community members.
In this article, we discuss several network analytic proxies that can be used to quantify different aspects of the social organisation in social collaboration networks.
We particularly focus on measures that can be related to the cohesiveness of the communities, the distribution of responsibilities and the resilience against turnover of community members.
We present a comparative analysis on a large-scale dataset that covers the full history of collaborations between users of $14$ major open source software communities.
Our analysis covers both aggregate and time-evolving measures and highlights differences in the social organisation across communities.
We argue that our results are a promising step towards the definition of suitable, potentially multi-dimensional, resilience and risk indicators for open source software communities.
\end{abstract}

\section{\label{intro}Introduction}

What are the most important social factors that lead to successful and sustainable open source software projects?
According to \emph{Linus' Law} - which states that ``given enough eyeballs, all bugs are shallow'' \cite{bazaar1999} - the quality and success of open source software (OSS) critically depends on the existence of a sufficiently large community of developers who review, modify and improve the publicly available source code.
Apart from development efforts, another important success factor is the existence of a stable community of users who report software defects, request and inspire new features, reproduce bugs or comment on issues reported by other users.
By employing the collective knowledge and diverse experiences of many contributors, most OSS communities manage to provide technical assistance to less experienced users, often on a time scale that is competitive to commercial software support.

Depending on the distribution of competencies and responsibilities of
contributors, largely different patterns of collaborations may arise.
While it is generally difficult to assess these social factors of OSS projects, the availability of large scale data on community dynamics increasingly allows to study the \emph{social dimension of OSS projects} from a quantitative perspective \cite{roblesturnover2006,mockuslongtermcontributions2012}.
Previous studies have mainly focused on rather simple proxies of social dynamics like the evolution of the number of contributors
and contributions or the time span of a user's activity and were mostly based on a rather limited set of snapshots of a single project.
Using a large scale dataset of time-stamped social interactions that has been collected from the \textsc{Bugzilla} bug-tracker installations of $14$ major OSS projects, in this paper we study the \emph{fine-grained evolution of structural features of networks of user collaborations}.
We thus take a \emph{network perspective on OSS communities} and highlight differences in the social organisation of software projects that can be related to their activity, their cohesion as well as their resilience against fluctuations in the community members.
By applying standard measures from social network analysis we particularly quantify how tightly community members collaborate, how equal responsibilities are distributed and how resilient collaboration topologies are against the loss of (central) community members.
While similar tools have been applied to OSS projects before
\cite{howisonsocial2006,niavaliditynet2010}, to the best of our
knowledge, the present paper is the first to study these network
analytic measures on a dataset that covers the full, fine-grained
history of $14$ well-established and successful OSS communities.

\section{\label{sec:methodology}Social Organisation in OSS
  Communities: A Network Perspective}

In order to make substantiated statements about the structure and dynamics of the social organisation of OSS communities, we recently completed collecting data on the history of user collaborations recorded by the \textsc{Bugzilla} installation of $14$ well-established OSS projects.
\textsc{Bugzilla}\cite{serranobugzilla2005} 
is an open source bug tracking system which is utilised by users and developers alike to report bugs, keep track of open issues and feature requests and comment on issues reported by others.
Since the \textsc{Bugzilla} installations of OSS projects are used to foster collaboration between community members, it constitutes a valuable source of data that allows us to track social interactions between developers and users.

\subsection{Building Social Networks from Bug-Reports}

Data in the \textsc{Bugzilla} database are arranged around the notion of \emph{bug reports}.
Each bug report has a set of fields describing aspects like the user
who initially filed the bug report, its current status
(e.g. \emph{pending}, \emph{reproduced}, \emph{solved}, etc)
, to whom the responsibility to provide a fix has been assigned, attachments which may be used to reproduce or resolve the issue, comments and hints by other community members, or a list of community members which shall be informed about future updates.
Apart from an initial bug report, \textsc{Bugzilla} additionally stores the full history of all updates to any of the fields of a bug report.
Each of these change records includes a time stamp, the ID of the user performing the change as well as the new values of the changed fields.
While our dataset comprises change records for all possible fields, in this article we focus on those that indicate changes in the users that are assigned responsibility to fix an issue (henceforth called the \emph{ASSIGNEE} field) and changes to the list of users to whom future updates of the bug shall be sent via E-Mail (henceforth called the \emph{CC} field).
We consider any updates in the \textit{CC} and \textit{ASSIGNEE} field of a bug report as a time-stamped edge from the user who performed the update to the user(s) who were added to the \textit{CC} field or the \textit{ASSIGNEE} list of responsible developers respectively.

Based on the data extraction procedure described above, we obtain a large time-aggregated network of nodes representing community members and time-stamped edges representing a particular interaction between two users.
For most of the projects considered, the \textsc{Bugzilla} history from which we extract the network is longer than ten years.
The fact that - in social networks aggregated over such long periods of time - most of the users represented by nodes have never been active within the same time period limits the expressiveness of the network structure in terms of a project's ``social organisation''.
In order to overcome this issue, we perform a \emph{dynamic network analysis} by defining a sequence of \emph{monthly collaboration networks} based on the time stamps of edges.
In particular, we define a 30 day sliding time window and filter out those edges whose time stamps are outside the window and those nodes who did not have any interactions in the corresponding time period.
By progressively advancing the start date of the sliding 30 day time window by one day increments we obtain a sequence of collaboration networks that allows us to study the structure of the community's social organisation as well as its evolution over time.
Naturally, most of the monthly networks obtained in the way described above will not be fully connected.
Since the network analytic measures we intend to apply assume connected topologies, we perform a component analysis on all snapshots and restrict our quantitative analysis to the largest connected component (LCC).
In order to test the significance of our findings we further compute the fraction of those nodes who are part of the largest connected component.
Table \ref{tab:bugs} shows the $14$ OSS projects that are included in our dataset along with the time period and the total number of bug reports and updates that we included in our analysis.
The column \emph{LCC/TOTAL} furthermore indicates the fraction of
users in the LCC, averaged over all monthly snapshots of the
corresponding project.
Here one observes that our data shows a rather large degree of variation with respect to this fraction, which may be seen as an argument that this measure is an interesting indicator for the \emph{cohesiveness} of OSS communities by itself.
Nevertheless, we argue that for all projects the fraction of users in the LCC is sufficiently large to make substantiated statements about the project's social organisation.
       
{
\begin{table}[htpb]
\setlength{\tabcolsep}{2pt}
\begin{center}
\caption{Aggregated measures for the studied projects. From column \emph{LCC/Total} to the last on the right, the numbers
  indicate the mean value $\pm$ standard deviation.
\label{tab:bugs}}
\scalebox{0.7}{
\begin{tabular}{c|c|c|c|c|c|c|c|c|c|c}
Project&\multirow{2}{*}{Bugs}&\multirow{2}{*}{Updates}&\multirow{2}{*}{Period}&\multirow{2}{*}{LCC/Total}&\multirow{2}{*}{Nodes
  in LCC}&\multirow{2}{*}{Edges}&Mean&\multirow{2}{*}{Assortativity}&Closeness&Clustering\\
Name&&&&&&&Degree&&Central.&Coefficient\\
\hline  
\rowcolor{Gray}\textsc{xamarin}&4552&20721&2011-2012&0.93$\pm$0.05&46.76$\pm$8.12&98.15$\pm$22.70&2.07$\pm$0.29&-0.14$\pm$0.11&0.40$\pm$0.07&0.22$\pm$0.05\\
\textsc{thunderbird}&35388&313957&2000-2012&0.53$\pm$0.26&64.82$\pm$53.49&86.44$\pm$80.05&1.05$\pm$0.42&-0.23$\pm$0.17&0.40$\pm$0.27&0.04$\pm$0.05\\
\rowcolor{Gray}\textsc{libreoffice}&8916&78341&2010-2012&0.78$\pm$0.11&73.83$\pm$32.06&114.41$\pm$49.10&1.56$\pm$0.26&-0.20$\pm$0.10&0.40$\pm$0.09&0.13$\pm$0.06\\
\textsc{mageia}&6600&46921&2006-2012&0.93$\pm$0.07&77.54$\pm$21.80&156.00$\pm$59.24&1.95$\pm$0.30&-0.37$\pm$0.12&0.54$\pm$0.09&0.14$\pm$0.04\\
\rowcolor{Gray}\textsc{mandriva}&60546&368463&2002-2012&0.70$\pm$0.18&88.15$\pm$60.70&142.16$\pm$118.44&1.41$\pm$0.38&-0.29$\pm$0.15&0.40$\pm$0.14&0.07$\pm$0.05\\
\textsc{firefox}&112953&1067914&1999-2012&0.58$\pm$0.23&171.77$\pm$117.79&240.79$\pm$180.44&1.16$\pm$0.44&-0.15$\pm$0.11&0.32$\pm$0.23&0.04$\pm$0.04\\
\rowcolor{Gray}\textsc{seamonkey}&90040&993392&1998-2012&0.67$\pm$0.15&210.39$\pm$251.43&364.42$\pm$482.54&1.48$\pm$0.48&-0.19$\pm$0.13&0.34$\pm$0.11&0.08$\pm$0.06\\
\textsc{netbeans}&210921&1875878&2000-2012&0.96$\pm$0.05&269.71$\pm$292.07&1069.72$\pm$1509.12&3.39$\pm$1.13&-0.12$\pm$0.08&0.37$\pm$0.05&0.23$\pm$0.08\\
\rowcolor{Gray}\textsc{openoffice}&118135&915749&2000-2012&0.88$\pm$0.19&319.01$\pm$169.88&931.35$\pm$591.80&2.52$\pm$0.84&-0.12$\pm$0.10&0.34$\pm$0.15&0.12$\pm$0.06\\
\textsc{gentoo}&140216&661783&2002-2012&0.80$\pm$0.07&338.97$\pm$110.86&617.73$\pm$211.92&1.82$\pm$0.27&-0.29$\pm$0.10&0.49$\pm$0.13&0.04$\pm$0.03\\
\rowcolor{Gray}\textsc{kde}&179470&648331&2002-2012&0.75$\pm$0.12&361.16$\pm$246.16&424.61$\pm$301.20&1.15$\pm$0.07&-0.16$\pm$0.07&0.32$\pm$0.07&0.01$\pm$0.01\\
\textsc{eclipse}&356415&2594385&2001-2012&0.78$\pm$0.08&472.58$\pm$180.71&964.47$\pm$411.94&2.06$\pm$0.38&0.05$\pm$0.08&0.25$\pm$0.05&0.13$\pm$0.03\\
\rowcolor{Gray}\textsc{gnome}&550722&2751441&2000-2012&0.67$\pm$0.12&523.76$\pm$585.26&610.16$\pm$616.81&1.25$\pm$0.22&-0.17$\pm$0.09&0.25$\pm$0.08&0.03$\pm$0.04\\
\textsc{redhat}&414163&3777634&2006-2012&0.45$\pm$0.26&658.06$\pm$865.97&983.58$\pm$1297.18&1.19$\pm$0.35&-0.12$\pm$0.20&0.30$\pm$0.23&0.00$\pm$0.01\\
\end{tabular}}  
\end{center} 
\end{table}   

\subsection{Network Measures}    
While the literature is rich in terms of measures able to quantify
structural features of networks \cite{stanleysocial1994,
  newmannetsbook2010}, due to space limitations here we focus on three
measures which are able to capture basic network qualities that relate
to the \emph{cohesiveness} of a community, the distribution of
responsibilities among its members and its resilience against fluctuations in the user base.
The first network measure is based on
the \emph{closeness centrality} of a node, which is defined as the
inverse of the sum of the shortest path length to all other nodes in
the network.   
\begin{equation}    
Cc(n_i)=\sum_{j=1, j\neq i}^N\frac{N-1}{d(n_i,n_j)} \in [0,1]
\label{eq:ccni}     
\end{equation}
where $Cc(n_i)$ corresponds to the \emph{closeness centrality} score
of node $n_i$, $d(n_i,n_j)$ is the length of the shortest path between
nodes $n_i$ and $n_j$, while $N$ corresponds to the total number of
nodes in a given network. Finally, the factor $N-1$ is a normalisation
constant \cite{freemancentrality1979}. Based on this, the \emph{closeness centralisation} of a
network $(Cc_{global})$ can be calculated by taking the sum of the differences between the
node with the highest value of closeness centrality $(n^*)$ and the closeness
centrality scores of all other nodes. This quantity is then normalised to the
range of $0$ to $1$ using the theoretical value that results from a
(maximally centralised) star network. Equation (\ref{eq:cc}) presents
the formal definition, while more details can be found in
\cite{freemancentrality1979,stanleysocial1994}. In the context of OSS collaboration networks, closeness centralisation
captures to what degree responsibilities, collaboration and
communication are distributed equally across community members. 

\begin{equation}
Cc_{global}=\sum_{i=1}^N\frac{Cc(n^*)-Cc(n_i)}{\frac{(N-2)(N-1)}{2N-3}}\in [0,1]
\label{eq:cc}
\end{equation}
     
The second measure, the \emph{clustering coefficient} of a network $(C)$,
measures how closely community members interact with each other in the
sense that an interaction between a user $X$ and $Y$, as well as an
interaction between user $Y$ and $Z$ will also entail a direct
interaction between the users $Y$ and $Z$. The formal definition is
presented in equations (\ref{eq:cni}) and (\ref{eq:c}).
 
\begin{equation}
C(n_i)=\frac{2 L_{D_{n_i}}}{D_{n_i}(D_{n_i}-1)} \in [0,1]
\label{eq:cni}
\end{equation}
\begin{equation}
C=\frac{1}{N}\sum_{i=1}^NC(n_i) \in [0,1]
\label{eq:c}
\end{equation}
where $D_{n_i}$ is the number of nodes directly connected to the node
$n_i$, while $L_{D_{n_i}}$ is the number of edges between
them. Therefore, the clustering coefficient $C(n_i)$ of node $n_i$
expresses the fraction of edges that were realised from the possible
$\frac{D_{n_i}(D_{n_i}-1)}{2}$ edges which are expected in a fully
connected network with $D_{n_i}$ nodes. We obtain the clustering
coefficient of a network by averaging the clustering
coefficient scores of all existing nodes (see equation (\ref{eq:c})). This procedure can be seen as measuring how
\emph{cohesive} the community is in terms of nodes being embedded in
collaborating clusters \cite{stanleysocial1994}.

Finally, the \emph{assortativity} $(r)$ measures an individual's preference to connect to other individuals that have a similar or different degree of connectivity (the degree being a node's number of connections to different nodes).
Networks in which nodes are preferentially connected to nodes with similar degree are called assortative.
In this case a positive degree assortativity $(0\ll r\leq1)$ indicates a positive correlation between the degrees of neighbouring nodes.
Networks in which nodes are preferentially connected to nodes with
different degree are called disassortative and in this case degree
assortativity is negative $(0\gg r\geq-1)$.
In networks with zero degree assortativity, there is no correlation
between the degrees of connected nodes, i.e. nodes do not exhibit a
preference for one or the other. Formally,
\begin{equation}
r=\frac{\sum_{ij}ij(e_{i,j}-q(i)q(j))}{\sigma(q)^2} \in [-1,1]
\label{eq:ass}
\end{equation}
where $e_{ij}$ is the fraction of all links in the network that join
together nodes with degrees $i$ and $j$, $q(i)=\sum_{j}e_{i,j}$, $q(j)=\sum_{i}e_{i,j}$ and $\sigma(q)$ is the standard deviation of the distribution of $q$.
The term $q(i)q(j)$ is the equivalent to the expected value of
$e_{i,j}$ inferred from a random network. Therefore, if $r=0$ the
pattern of interconnection between nodes is also random \cite{newmanR2003}.
 
\section{\label{sec:comp}Comparative Analysis of OSS Communities}
  
As described above, the preliminary results presented here have been obtained for the LCC of the network of monthly collaborations in terms of \textit{CC} and \textit{ASSIGNEE} interactions. While Table \ref{tab:bugs} shows the aggregate measures averaged over all time windows for every project in our database, due to space constraints we limit the presentation of the dynamics of the social organisation to the projects \textsc{Gentoo} and \textsc{KDE} (both \textsc{Gnu/Linux} related projects) as well as \textsc{Eclipse} and \textsc{NetBeans} (both \textsc{Java} IDEs).
These have been chosen because a) their communities are of comparable
size and age, b) the respective pairs of projects address similar
problem domains and c) they represent contrasting examples with
respect to the measures studied in this paper.

Figure \ref{fig:time_results} shows the evolution of the number of nodes in the LCC, its assortativity, clustering coefficient and closeness centralisation for these four projects.
For all projects, the fraction of nodes in the LCC is rather stable with values between $0.7$ and $1$ consistent with the aggregate values given in Table \ref{tab:bugs}.
The same is true for the evolution of the mean degree.
We thus omit these plots. The four projects show significant differences in the evolution of the clustering coefficient that cannot be explained by mere size effects.
In the particular time frame between $2006$ and $2008$, the clustering coefficient of the \textsc{Eclipse} community ($\approx 0.15$) was roughly ten times higher than that of the \textsc{Gentoo} community ($\approx 0.01$), although the LCCs of both communities were of comparable size ($\approx 500$ nodes).
In addition, the clustering coefficient of the \textsc{Gentoo}
community shows an interesting dynamics, dropping to a very small
value between $2006$ and $2008$ and increasing
thereafter. 

\begin{figure}[htpb]
\centering
    \subfigure[Nodes in LCC\label{fig:gentookde:nodes}]{
      \includegraphics[width=3.1cm]{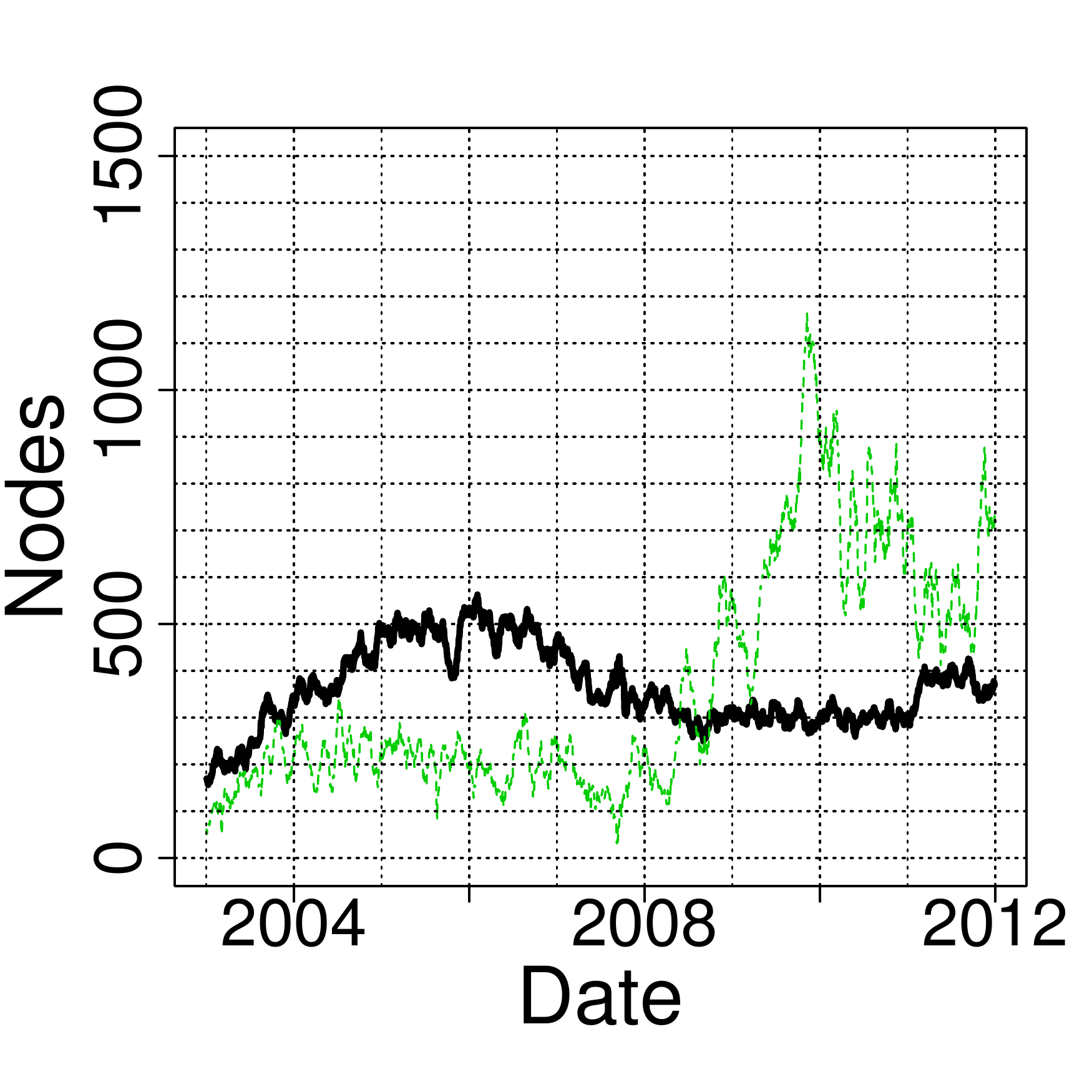}
    }
    \subfigure[Assortativity\label{fig:gentookde:ass}]{
      \includegraphics[width=3.1cm]{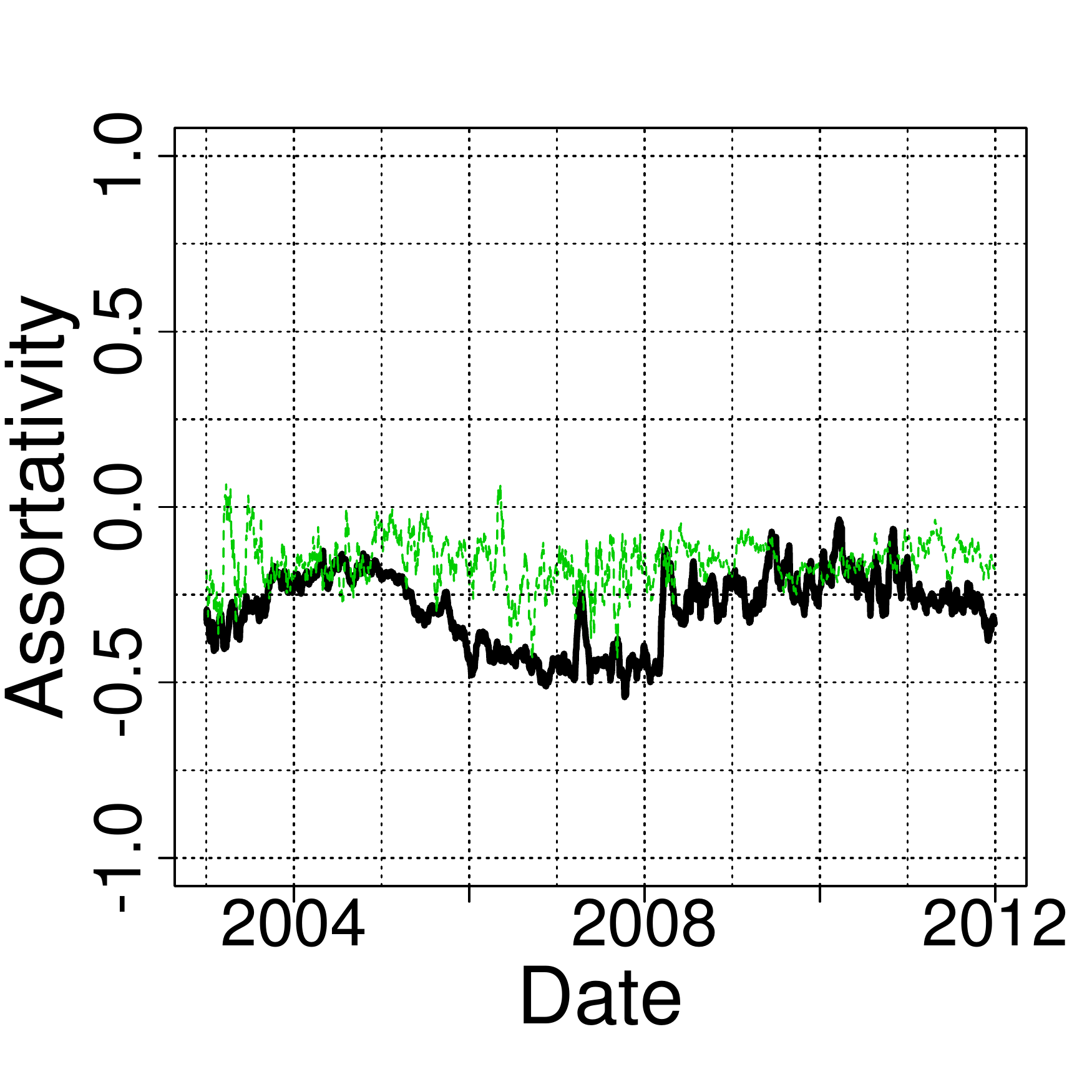}
    }
    \subfigure[Clustering coeff.\label{fig:gentookde:clucoe}]{
      \includegraphics[width=3.1cm]{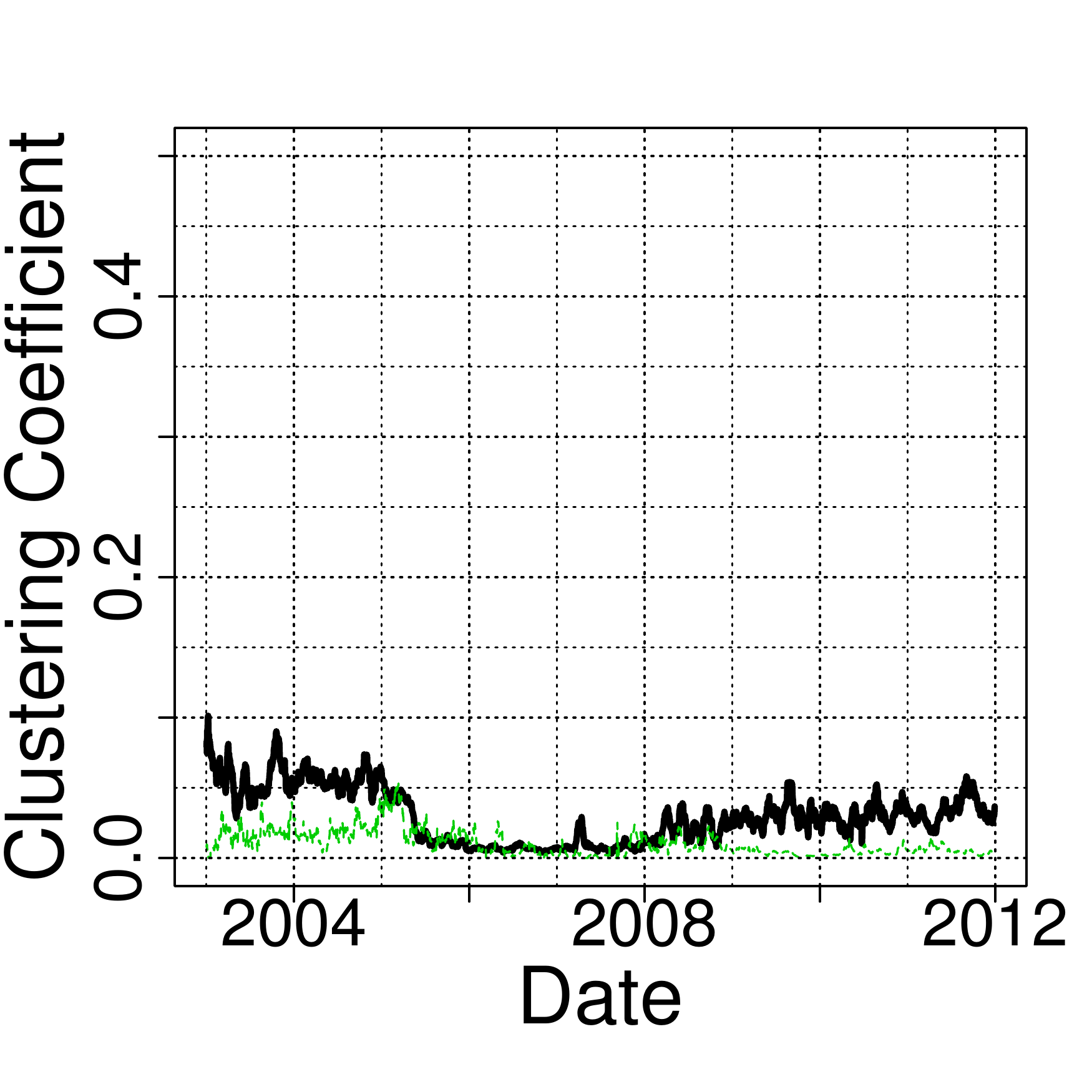}
    }
    \subfigure[Closeness central.\label{fig:gentookde:clocent}]{
      \includegraphics[width=3.1cm]{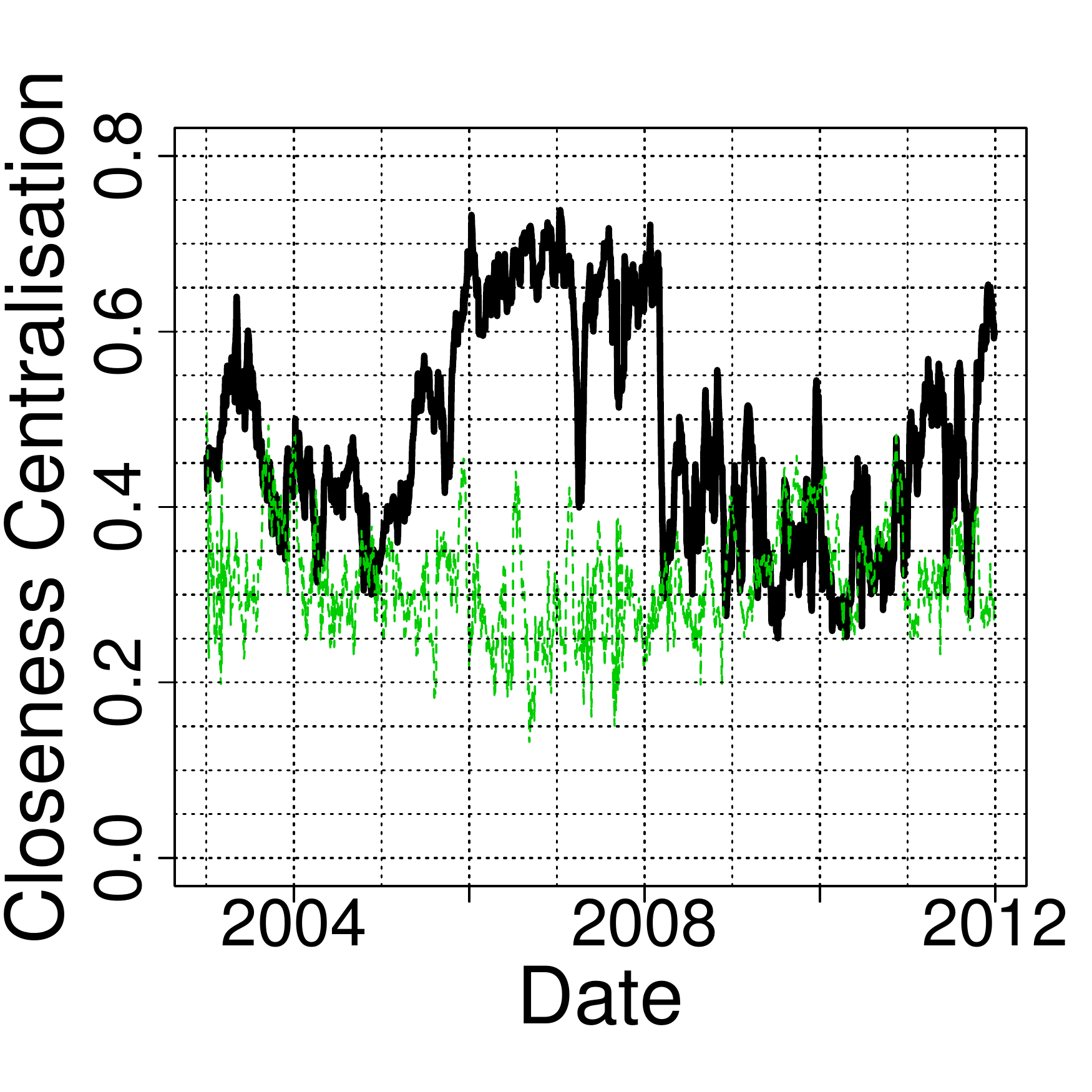}
    }
    \subfigure[Nodes in LCC\label{fig:eclipse_netbeans:nodes}]{
      \includegraphics[width=3.1cm]{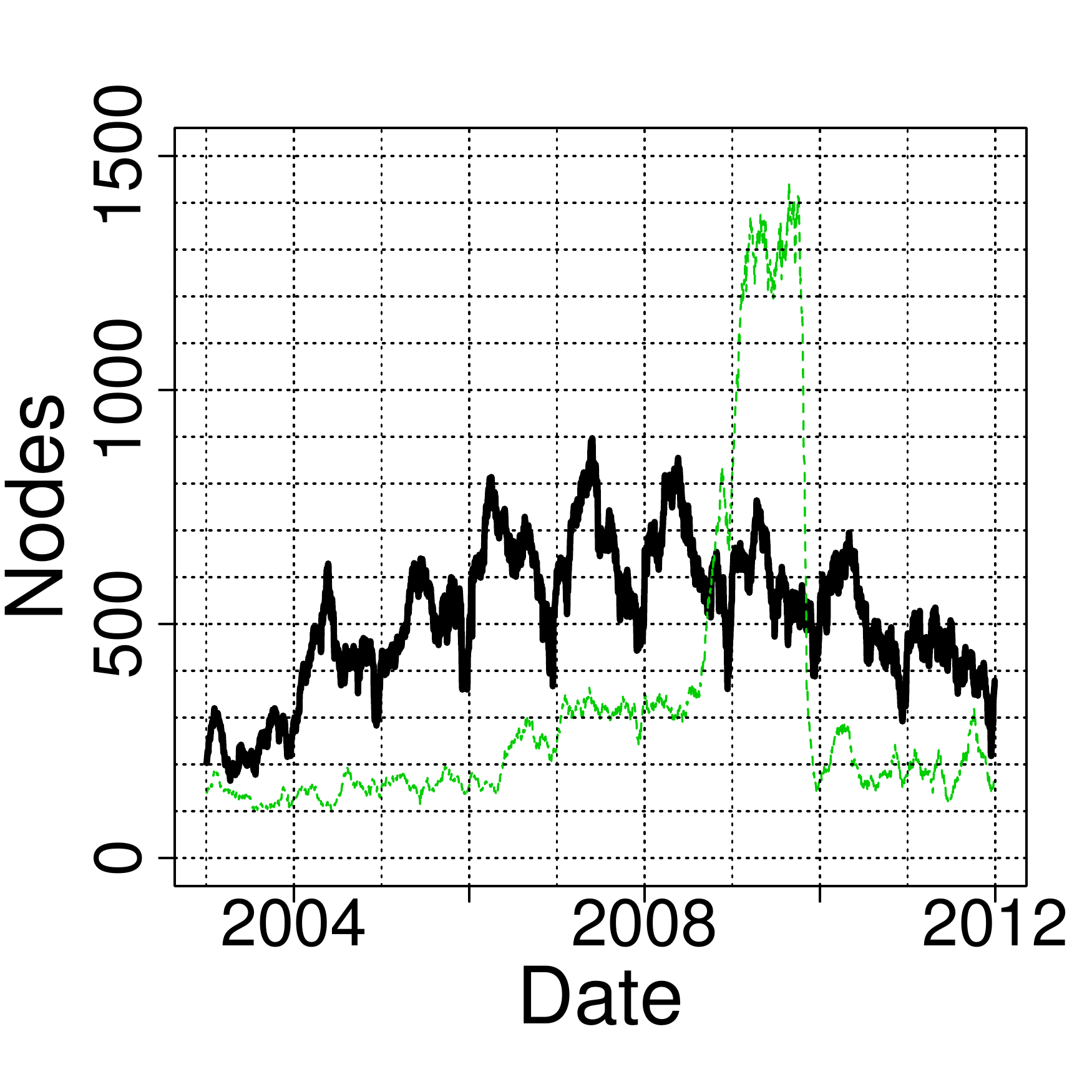}
    }
    \subfigure[Assortativity\label{fig:eclipse_netbeans:ass}]{
      \includegraphics[width=3.1cm]{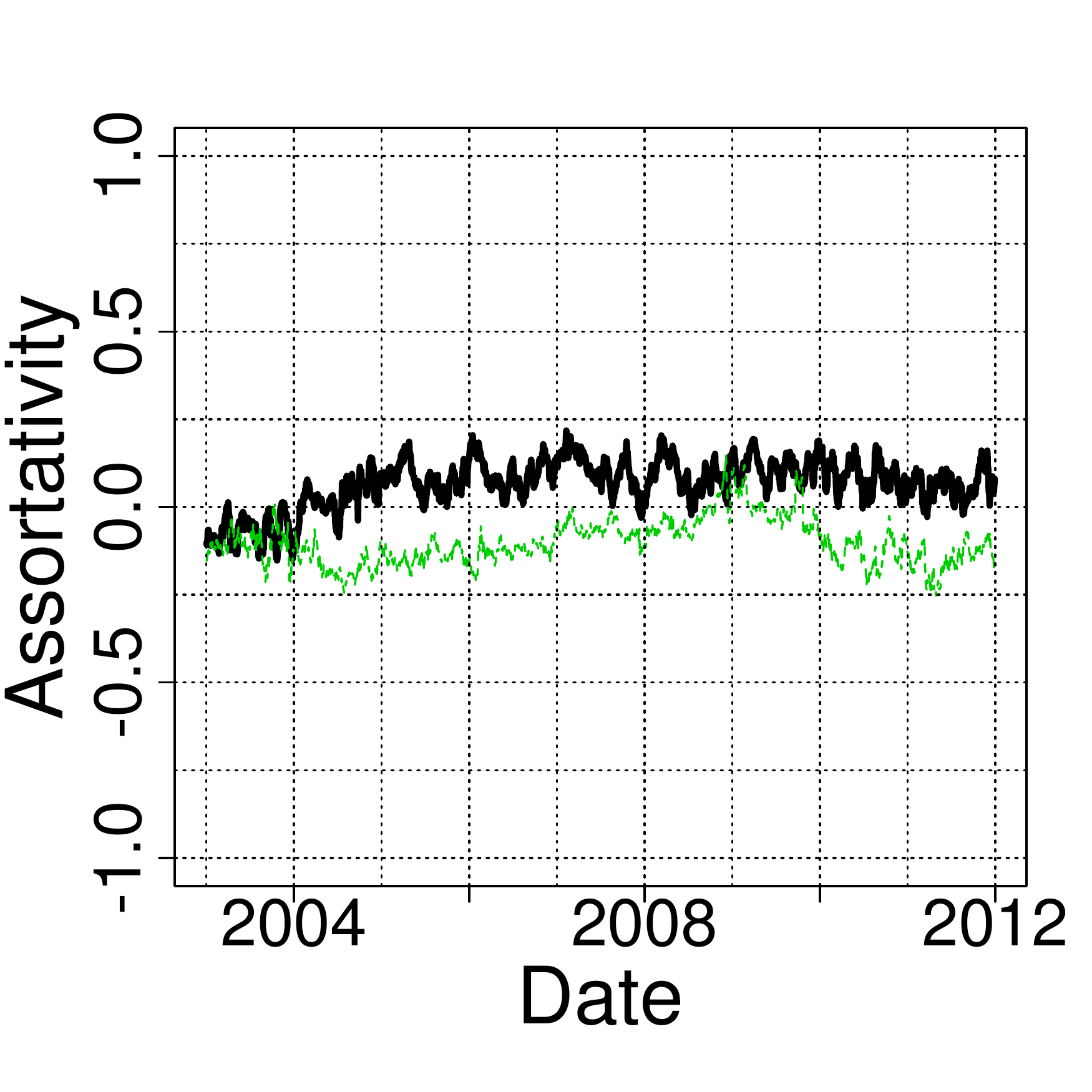}
    }
    \subfigure[Clustering coeff.\label{fig:eclipse_netbeans:clucoe}]{
      \includegraphics[width=3.1cm]{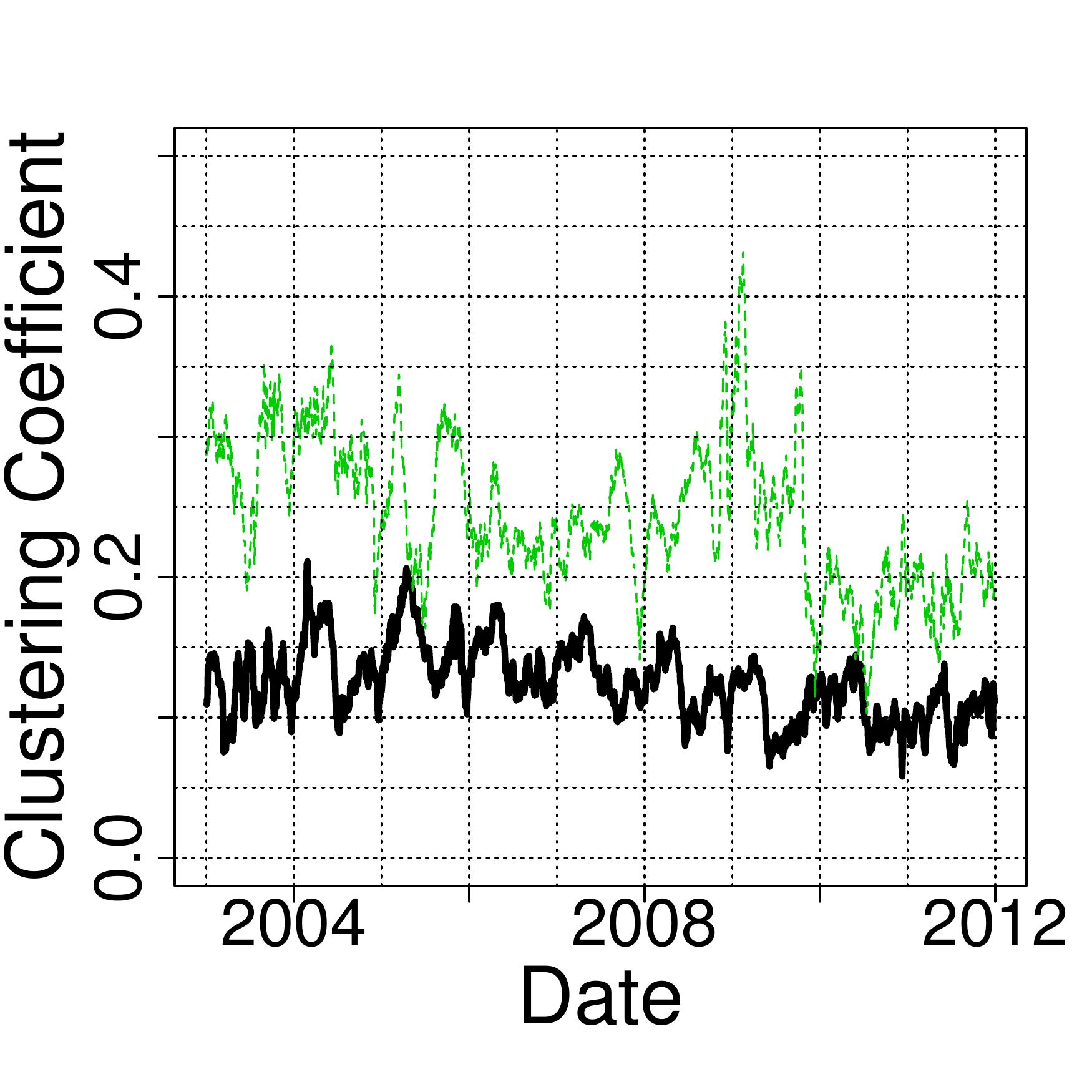}
    }
    \subfigure[Closeness central.\label{fig:eclipse_netbeans:clocent}]{
      \includegraphics[width=3.1cm]{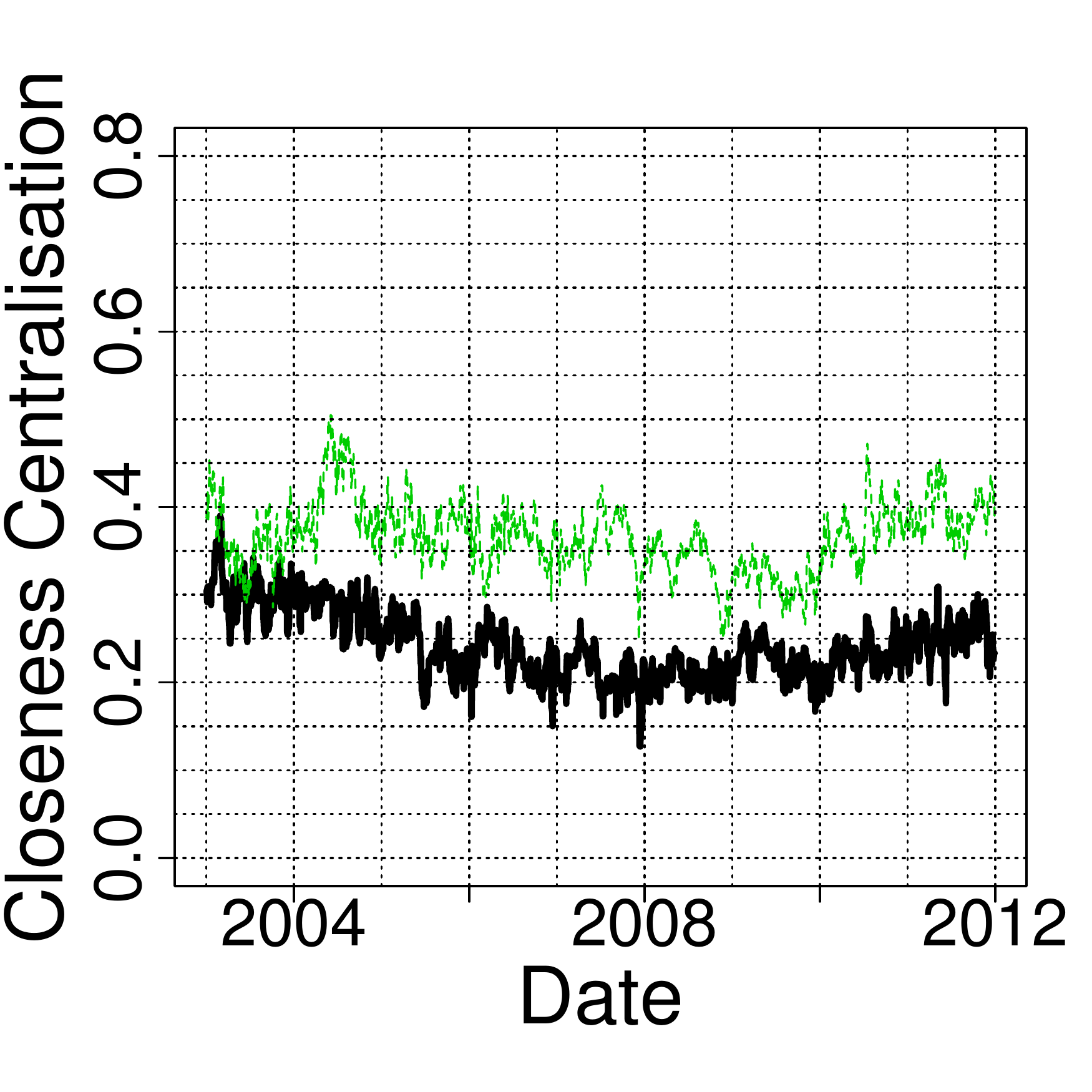}
    } 
\caption{Evolution of structural measures of the LCC in the monthly
  \textsc{Bugzilla} collaboration networks. (a-d): \textsc{Gnu/Linux}
  related projects \textsc{Gentoo} (black) and \textsc{KDE} (green),
  (e-h): IDEs \textsc{Eclipse} (black) and \textsc{NetBeans} (green)\label{fig:time_results}.} 
\end{figure}

\begin{figure}[htpb]
\centering
    \subfigure[\textsc{Gentoo} (Jan/2006) \newline $nodes=535, edges=785$\label{fig:gentoo_net}]{
      \includegraphics[width=0.22\textwidth]{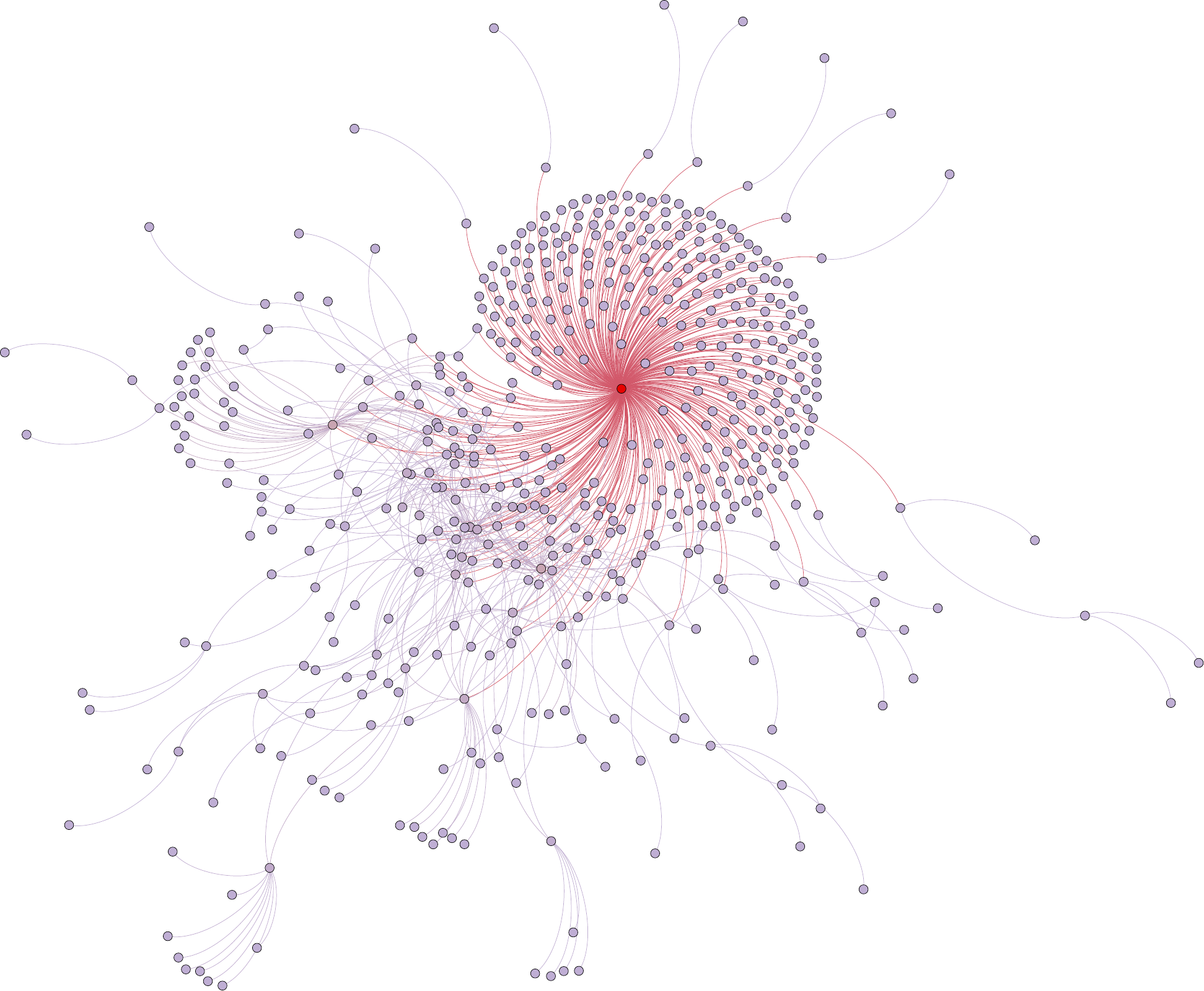}
    }
    \subfigure[\textsc{KDE} (Feb/2011) \newline $nodes=543, edges=630$\label{fig:kde_net}]{
      \includegraphics[width=0.22\textwidth]{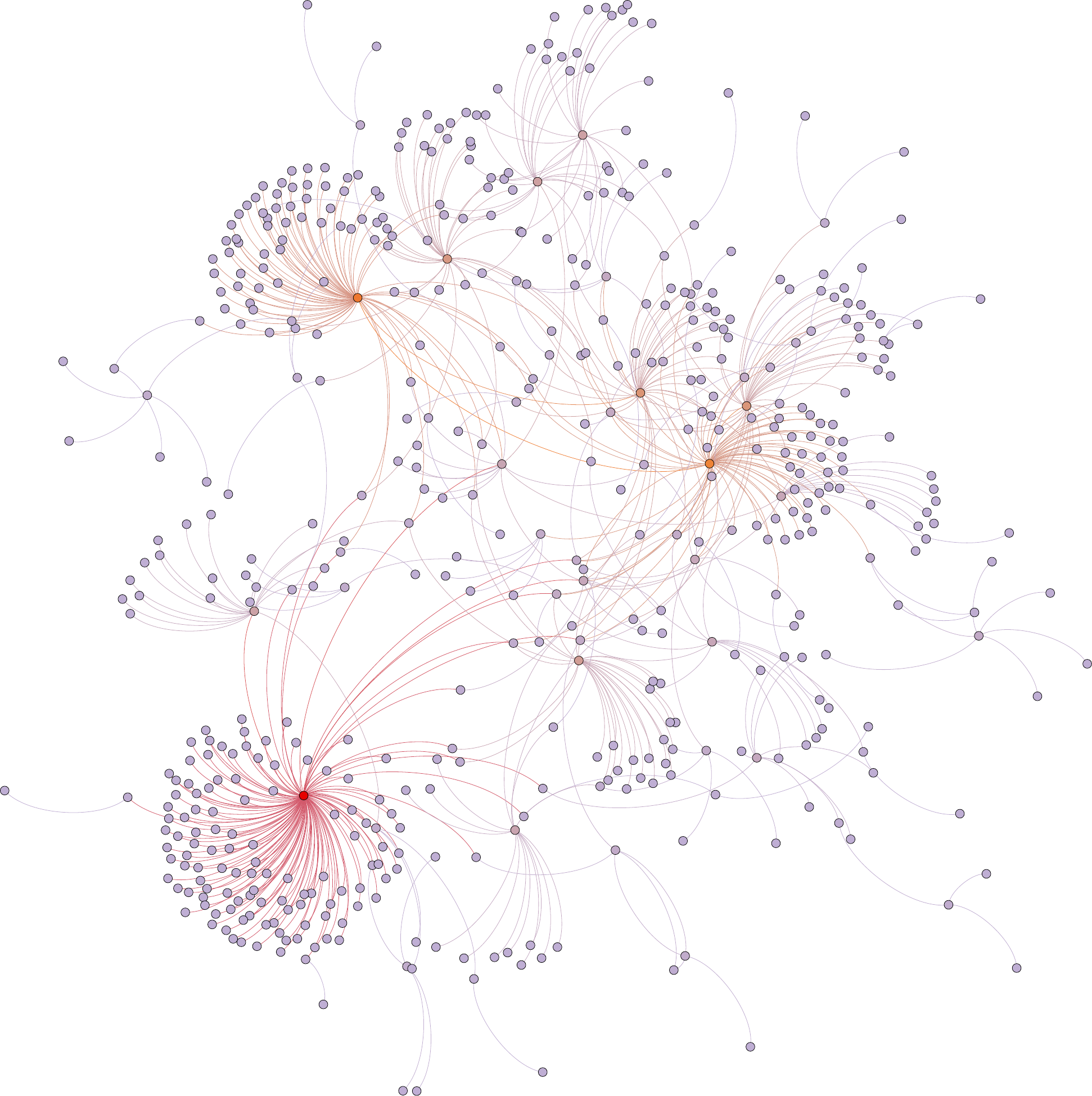}
    }
    \subfigure[\textsc{Eclipse} (Jan/2010) \newline $nodes=502, edges=868$\label{fig:eclipse_net}]{
      \includegraphics[width=0.22\textwidth]{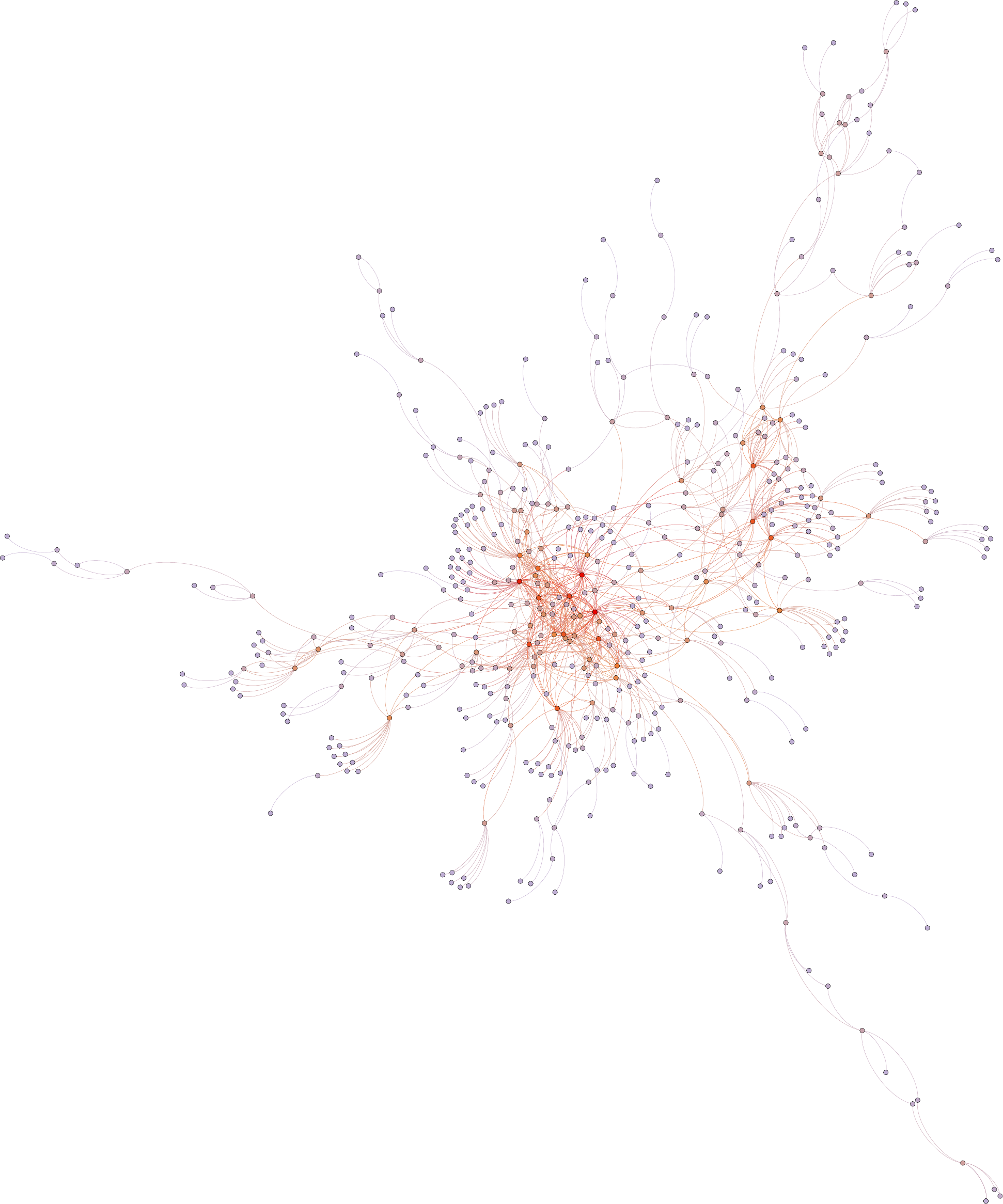}
    }
\subfigure[\textsc{Netbeans} (Sep/2008) \newline $nodes=566, edges=2753$\label{fig:netbeans_net}]{
      \includegraphics[width=0.22\textwidth]{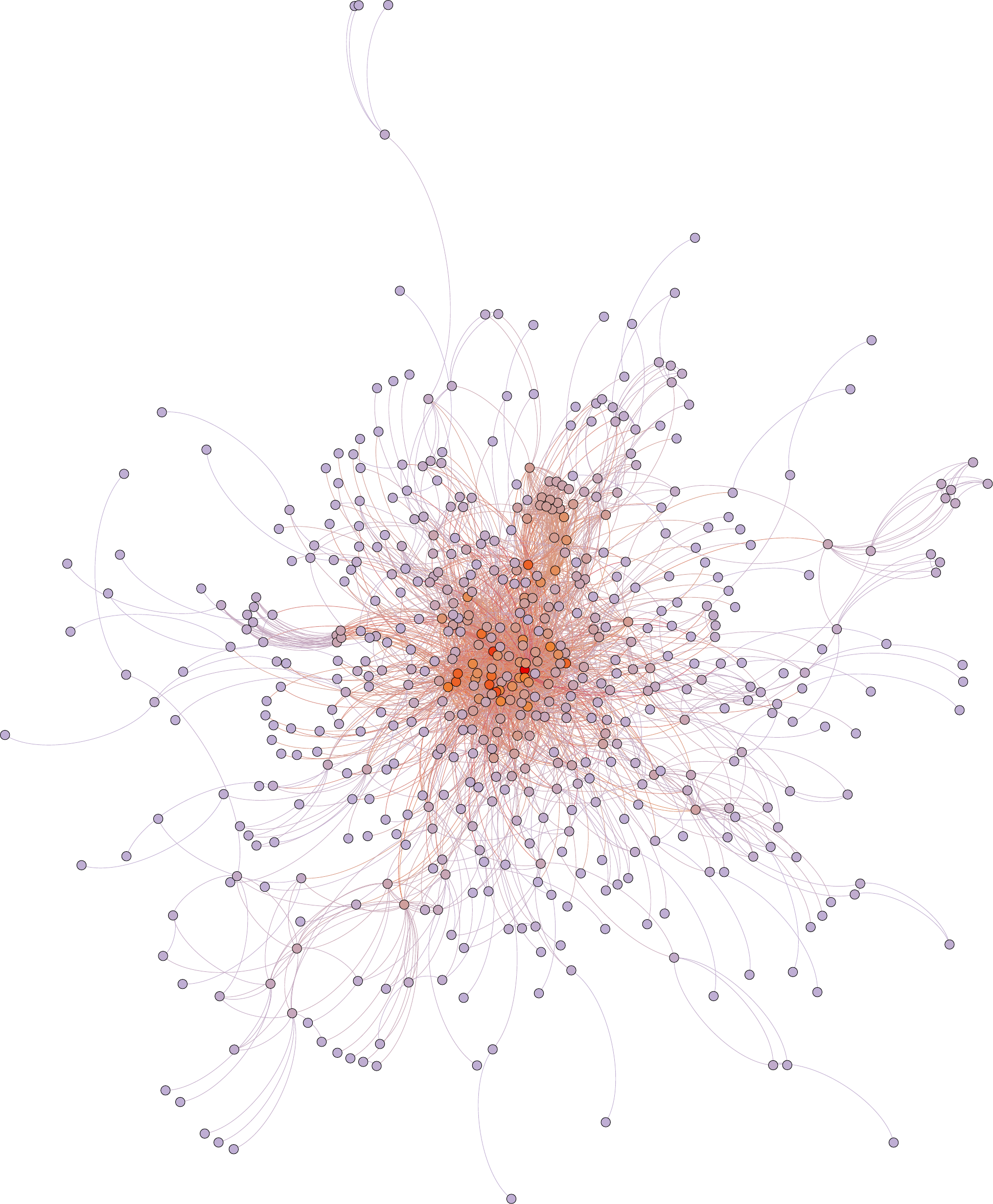}
    }
    \caption{Four monthly collaboration networks with comparable
      size showing largely different social
      organisation\label{fig:networks} (the visualisation was
      generated by \textsc{Gephi} \cite{gephi}).}
\end{figure}   
   
A different perspective of the structural change the \textsc{Gentoo} community was undergoing is given in Figure \ref{fig:gentookde:clocent} which displays a visible plateau in the closeness centralisation of the network within the same period.
In fact, as can be seen in the network depicted in Figure
\ref{fig:gentoo_net}, in the period between $2006$ and $2008$ most of the
collaborations were mediated by a single central community member,
while the social organisation of the \textsc{Eclipse} community
depicted in \ref{fig:eclipse_net} was structured in a much more
homogeneous way. 
The evolution of degree assortativity is captured in Figures \ref{fig:gentookde:ass} and \ref{fig:eclipse_netbeans:ass}.
Both the level of degree assortativity as well as its dynamics differ across the projects.
The collaboration network of \textsc{Eclipse}\textsc{} exhibits a
tendency towards assortative structures (meaning that high degree
nodes are preferentially connected to high degree nodes).  
The opposite is true for the \textsc{KDE} and the \textsc{Gentoo} communities which show a tendency towards disassortativity.
We thus argue that assortativity is suitable to further differentiate the social organisation of OSS communities.
%
%

 
\section{\label{con}Conclusions and Future Work}
We have studied measures that capture different structural dimensions in the social organisation of OSS projects.
Our analysis is based on a comprehensive dataset collected from the bug tracking communities of $14$ major OSS projects.
We view the social organisation from the perspective of time-evolving networks and highlight how projects, although similar in terms of size, problem domain and age, a) largely differ in terms of clustering coefficient, assortativity and closeness centralisation and b) that some projects show interesting dynamics with respect to these measures that cannot be explained by mere size effects.
We argue that the phase of high closeness centralisation and low clustering coefficient observed in the \textsc{Gentoo} community between $2006$ and $2008$ may be interpreted as a lack of \emph{social cohesion} which can possibly pose a risk for the project.

While our results are necessarily preliminary, we
currently extend our work by adding spectral measures like algebraic
connectivity and inequality measures like the Gini index that can
highlight further differences in the social organisation
\cite{zanettiACS2013}. A detailed case study is under
preparation \cite{zanettiICSE2013} and further includes community
performance indicators (e.g. response times, bug fixing times and
fraction of open issues) that can be mined from our dataset. The eventual goal of our project is the provision of multi-dimensional indicators for the social and technical organisation of OSS projects that are correlated with performance and that can be considered in the management and evaluation of OSS projects \cite{zanettiDOCSYMPICSE2012,zanettiQmodularity12,sustainablegrowthtessone2011}.
Such indicators can be useful when taking informed decisions
about which OSS project to invest in or rely on. Furthermore, due to
the distributed nature of collaborations, individuals often lack a global perspective on evolving communication and coordination
structures, even though these can influence long-term success. An inclusion of suitable indicators in community platforms like e.g. \textsc{Bugzilla} can assist in determining risks and allow project managers to timely react by shifting responsibilities, fostering information flow or changing organisational procedures.

\section{Acknowledgement}

This work was supported by the SNF through grant CR12I1\_125298. C.J.T. acknowledges financial
support from the SBF through grant C09.0055.
 
\bibliography{ref}
  
\end{document}